\documentclass[journal]{IEEEtran}
\usepackage{mathrsfs}
\usepackage{amsmath}
\usepackage{mathrsfs}
\usepackage{stmaryrd}
\usepackage{amsfonts}
\usepackage{algorithm}
\usepackage{algorithmic}
\usepackage{subfigure}
\usepackage{cite}
\usepackage{blindtext}
\usepackage{amssymb}
\usepackage{esint}
\usepackage{graphicx,xcolor}
\usepackage[]{graphicx}
\usepackage{epstopdf}
\usepackage{epsfig}
\usepackage{color}
\usepackage{subfig}
\usepackage[section] {placeins} 
\usepackage{morefloats} 

\newtheorem{theorem}{Theorem}
\newtheorem{remark}{Remark}[section]
\newcommand{\dkh}[1]{\left\{#1\right\}}
\newcommand{\norm}[1]{\left\Vert #1 \right\Vert}
\newcommand{\be}{\begin{eqnarray}}
\newcommand{\ee}{\end{eqnarray}}
\newcommand{\ce}{\begin{eqnarray*}}
\newcommand{\de}{\end{eqnarray*}}
\def\Rp{\mathbb{R}^{p}}

\def\Rp{\mathbb{R}^{p}}

\ifCLASSINFOpdf
\else
\fi
\begin{document}
%
\title{A Primal Dual Active Set Algorithm with Continuation for Compressed Sensing}

\author{Qibin~Fan,
        Yuling~Jiao,
         Xiliang Lu
\thanks{Qibin Fan, Yuling~Jiao and Xiliang Lu (Corresponding author) are in the School of Mathematics and Statistics, Wuhan University, Wuhan, China.
e-mails: qbfan@whu.edu.cn; yulingjiaomath@whu.edu.cn; xllv.math@whu.edu.cn}
}

\markboth{ IEEE TRANSECTION ON SIGNAL PROCESSING,Vol.~~, No.~~, ~~,2013}%
{Shell \MakeLowercase{\textit{et al.}}: Bare Demo of IEEEtran.cls for Journals}
\maketitle

\begin{abstract}
The success of compressed sensing relies essentially on the ability to efficiently find
an approximately sparse solution to an under-determined linear system. In this paper, we developed
an efficient algorithm for the sparsity promoting $\ell_1$-regularized least squares problem
by coupling the primal dual active set strategy with a continuation technique (on the regularization
parameter). In the active set strategy, we first determine the active set from
primal and dual variables, and then update the primal and dual variables by solving a
low-dimensional least square problem on the active set, which makes the algorithm very efficient.
The continuation technique globalizes the convergence of the algorithm, with provable global convergence
under restricted isometry property (RIP). Further, we adopt two alternative methods, i.e., a modified discrepancy
principle and a Bayesian information criterion, to choose the regularization parameter. Numerical
experiments indicate that our algorithm is very competitive with state-of-the-art algorithms in
terms of accuracy and efficiency.
\end{abstract}

%
\IEEEpeerreviewmaketitle

\section{Introduction}
\IEEEPARstart{C}{ompressive} sensing (CS) has recently emerged as a promising approach for acquiring (approximately) sparse
signals. An important problem in CS is to find the sparsest solution of the following under-determined
linear system \cite{cs1,cs2,cs3}
\begin{equation}\label{equ:gov}
\Psi x = y, 
\end{equation}
where $\Psi  \in \mathbb{R}^{n\times p} $ is the sampling matrix with $n\ll p$, $x$ is a sparse
signal, $y$ is the measurement, which may contain noise. It can be equivalently written as an optimization problem
\begin{equation}\label{l0noise}
 \min_{x \in \mathbb{R}^{p}}\|x\|_{0} ,   \quad \textrm{subject }  \quad \textrm{to }  \quad \|\Psi x
 -y\|_{2}\leq \epsilon,
\end{equation}
where $\|x\|_{0}$ denotes the the number of nonzero entries in the vector $x$ and $\epsilon$ is the noise level.
Due to the nonsmooth and nonconvex structure of problem \eqref{l0noise}, it is very challenging to find
the sparsest solution. Now it is widely accepted that the $\ell_1$ convex
relaxation can provide a satisfactory approximate solution, if the solution $x$ and
the sampling matrix $\Psi$ satisfies certain conditions.

There are three different versions of $\ell_1$ convex relaxation that have received a lot of attentions. They are Basis Pursuit Denoising (BPDN) \cite{bp}:
\begin{equation}\label{bpd}
 \min_{x \in \mathbb{R}^{p}}\|x\|_{1} ,   \quad \textrm{subject }  \quad \textrm{to }  \quad \|\Psi x
 -y\|_{2}\leq \epsilon,
\end{equation}
the $\ell_1$-regularized least squares problem \cite{bp}:
\begin{equation}\label{rgl1*}
 \min_{x \in \mathbb{R}^{p}}  \tfrac{1}{2}\|\Psi x
 -y\|^{2}_{2} + \lambda\|x\|_{1} ,
\end{equation}
and the least absolute shrinkage and selection operator (LASSO) \cite{lasso} model:
\begin{equation}\label{lasso}
 \min_{x \in \mathbb{R}^{p}}\|\Psi x
 -y\|_{2},   \quad \textrm{subject }  \quad \textrm{to }  \quad \|x\|_{1} \leq\tau.
\end{equation}
Where  $\lambda$ and $\tau$ are regularized parameter and tuning  parameter, respectively.
It was shown in \cite{pareto} if these parameters are chosen properly, problems (\ref{bpd}) - (\ref{lasso}) have the same minimizer. In this paper, we are interested in the fast solution of the $\ell_1$-regularized least squares model \eqref{rgl1*}.

Over the last few years, a large number of algorithms have been developed
for problems (\ref{bpd}) - (\ref{lasso}). We will list only a few exemplary
methods here, and refer to the review \cite{csnmr,BJMG,pfbsr} for a comprehensive overview.
Gradient type methods, e.g., gradient projection sparse
reconstruction  \cite{gpsr}, sparse reconstruction via separable
approximation \cite{sparsa}, spectral gradient projection \cite{pareto},
fixed point iteration with continuation strategy \cite{fpc,wen2010fast}, iterative shrinking/thresholding
algorithm \cite{ista, pfbs} and their accelerated extension  \cite{nest2},\cite{fista}, \cite{nesta}, are extremely
popular. Other classical methods, e.g., homotopy method \cite{homtop1,homtop2,homtop3}, alternating direction
method of multipliers \cite{admr}, iteratively reweighted least square method \cite{irls1},
have also received revived interest in solving $\ell_1$ minimization problems.

These algorithms can have only  sublinear or linear convergence rate.
Therefore, it is of immense interest to develop Newton type algorithms that enjoy a (locally)
superlinear convergence rate. For an invertible matrix $\Psi$, the primal dual
active set (PDAS) method (also known as semismooth Newton method), has been studied in \cite{ssn1,ssn3,ssn4}.
This idea can be extended to the CS setting to solve problem \eqref{rgl1*}.
Theoretically, it enjoys a locally superlinear convergence.

However, in  Newton type algorithm, a good initial guess is very important for the successful application
of the PDAS method. Unlike gradient based algorithms, the PDAS method does not have a monotonic decreasing
property for the cost functional. Therefore without a good initial guess the algorithm may not converge.
Meanwhile, in the model \eqref{rgl1*}, the regularization parameter $\lambda $ balances the sparsity of the
solution and the fidelity of the measurements. And its proper choice plays an essential role for getting
a satisfactory reconstruction.

In this article we propose a simple but efficient technique to find a good initial guess by combining the continuation strategy with primal dual active set algorithm. Moreover, equipped with a proper stop rule, the regularization parameter can be chosen automatically without much adding work. To be precise, the $\ell_1$-regularized minimization problems
are solved with warm start on a predefined decreasing sequence $\{\lambda_s\}_s$,  i.e,  the solution $x_({\lambda_s})$  to
$\lambda_s$-problem is chosen as the initial guess for $\lambda_{s+1}$-problem solved by PDAS. It needs only a few
(Newton) steps since $x({\lambda_s})$ provides a good initial guess.

The main contributions of this paper are twofold. First, we derive a local one step convergence result for $\lambda_{s+1}$-problem which improves the well known local supperliner convergence of PDAS \cite{ssn1}. More importantly, we prove
the global convergence for the primal dual active set algorithm with continuation (PDASC) under the standard  restricted isometry property (RIP) assumption on the matrix $\Psi$ in the noise-free case. On the other hand, when the measurement involves noise, we adopt the parameter selection rule based on either a modified discrepancy principle
or Bayesian information criterion. One can use this rule to select a suitable regularization parameter $\hat{\lambda}$ and solution $x(\hat{\lambda})$  during the continuation process with nearly no adding effort.


The rest of the paper is organized as follows. In section 2 we introduce
the mathematical background, PDAS algorithm and continuation technique,  discuss their
convergence properties and the  regularization parameter selection rule. In section 3,
several numerical examples are presented to illustrate the efficiency and accuracy PDASC algorithm,
by comparing  with several state-of-the-art sparse reconstruction algorithms.
The technical proofs are in the appendices.

\section{PDASC Algorithm}
\subsection{Notations}
Given a vector $x = (x_{1},x_{2},...,x_{p})^t \in \mathbb{R}^{p}$, we denote
by $\|x\|_q = (\sum_{i=1}^{p}|x_{i}|^q)^\frac{1}{q}$ with $q\in [1,\infty)$ and
$\|x\|_{\infty} = \max_{1\leq i \leq p} |x_i|$. Further,
$\Psi^t$ and $\norm \Psi$ denote the transpose and 2-norm of the matrix $\Psi$, respectively. The matrix $\Psi$ is assumed to be columnwise normalized, i.e., $\|\Psi_{i}\|_2 = 1$ for $i=1,...,p$.
The notation $\textbf{1}$ (or $\textbf{0}$) refers to a column vector with all entries equal to 1 (or 0).
For any set
$$ A \subseteq S \triangleq \{1,2,...,p\}$$
of size $|A|$,
$x_{A}\in \mathbb{R}^{|A|}(\Psi_{A}\in \mathbb{R}^{n\times|A|})$ is the
subvector (submatrix) whose entries (columns) are listed in $A$.

We denote by $\Gamma_{0}(\mathbb{R}^{p})$
the set of all proper lower semicontinuous convex functions on $\mathbb{R}^{p}$.
The subdifferential of any $f\in \Gamma_{0}(\mathbb{R}^{p})$ is a set-value mapping defined by
\begin{equation*}
 \partial f (z):= \{w\in \mathbb{R}^{p}:f(v)\geq f(z)+ \langle w,v-z\rangle, \forall v \in \mathbb{R}^{p}\}.
\end{equation*}
The subdifferential of $f=\|x\|_{1}$ is the pointwise
set-value sign function $\mathrm{Sign}(x)$ \cite{orgth}, i.e.,
 \begin{equation}\label{subdiff}
 z \in  \mathrm{Sign}(x) \Leftrightarrow z_{i}
  \left\{
    \begin{array}{ll}
   $= 1$,    \quad &\text{$x_{i}>0,$}\\
   $= -1$  ,  \quad &\text{$x_{i}<0,$}\\
   $$\in$ [-1,1]$,  \quad &\text{$x_{i} = 0.$}
    \end{array}
  \right.
\end{equation}
The classical Fermat's rule for proper lower semicontinuous convex functions \cite{roc} asserts
\begin{equation}\label{fermat}
\textbf{0} \in \partial f(z^{*}) \Leftrightarrow z^{*} \,\, \text{is a minimizer of} \,\, f.
\end{equation}
For a given $f\in\Gamma_0(\mathbb{R}^p)$, the proximal operator $Prox_{f}$ is defined by
$
Prox_{f}(z) := \mathop \text{argmin}_{x\in \mathbb{R}^{p}} \{\tfrac{1}{2}\|{x-z}\|^2 + f(x)\}.
$
Then there holds \cite{general}
\begin{equation}\label{general}
w \in  \partial   f(z) \Leftrightarrow z= Prox_{f}(z+w).
\end{equation}
The proximal operator of $\|\cdot\|_{1}$ is given
by the pointwise soft-thresholding operator  \cite{orgth}
\begin{equation}\label{proxl1}
Prox_{\lambda \| x\|_{1}}(z) = T_{\lambda }(x),
\end{equation}
where
\begin{equation}\label{softth}
z=T_{\lambda}(x)  \Leftrightarrow z_{i}=\max\{|x_{i}|-\lambda,0\}\mathrm{sign}(x_{i}).
\end{equation}

\subsection{Motivation and PDAS Algorithm}
Now we characterize the minimizer of (\ref{rgl1*}) by its KKT system (c.f. \cite{fpc}), which motivates the  PDAS algorithm; see
also Appendix A for a short proof, which is included for completeness.

\begin{theorem}\label{th1}
If $x^{*} \in \Rp$ is a minimizer of (\ref{rgl1*}), then there exists a $d^{*} \in \Rp$ such that the
\textrm{KKT} system  holds:
\begin{eqnarray}
\Psi^t \Psi x^{*} + d^{*} = \Psi ^{t} y,  \label{K1} \\
x^{*} = T_{\lambda}(x^{*} + d^{*}).  \label{K2}
\end{eqnarray}
Conversely, if $x^{*}\in \Rp$ and $d^{*} \in \Rp$ satisfying (\ref{K1}) and (\ref{K2}),
then $x^{*}$ is a minimizer of (\ref{rgl1*}).
\end{theorem}

Let $x^*$ and $d^*$ be the optimal primal and dual variables. Clearly, it follows from  (\ref{K2}) that
\begin{equation*}
x^*_i > 0 \Leftrightarrow  x^{*}_{i} + d^{*}_{i} > \lambda, \quad x^*_i < 0 \Leftrightarrow  x^{*}_{i} + d^{*}_{i} < -\lambda.
\end{equation*}
Hence, one can use the information from both primal and dual variables, rather than the primal
variable alone, to determine the nonzero components of $x_i^*$ (which is called active set).
This motivates us to define the active and inactive sets by:
\begin{equation}\label{equ:active}
\left. \begin{array}{l}A_{*}^{+} = \dkh{i \in S: x^{*}_{i} + d^{*}_{i} > \lambda},\\
A_{*}^{-} = \dkh{i \in S: x^{*}_{i} + d^{*}_{i} <- \lambda},\\
A_{*} = A_{*}^{+} \cup A_{*}^{-}, \quad I_{*} = A_{*}^{c}.
\end{array}\right\}
\end{equation}
Then the KKT system (\ref{K1})-(\ref{K2}) can be reformulated.
First, by (\ref{K2}) and the soft thresholding
operator (\ref{softth}), we deduce
\begin{equation}\label{reducer1}
x_{I_{*}}^{*} = \textbf{0}_{I_{*}}^{*}.
\end{equation}
Meanwhile, the proof of Theorem \ref{th1} implies $d^{*} \in \lambda\partial\|\cdot\|_1(x^{*})$. Then by (\ref{subdiff}), we get
$d_{A_{*}^{+}}^{*} = \lambda \textbf{1}_{A_{*}^{+}}$, and $d_{A_{*}^{-}}^{*} = -\lambda\textbf{1}_{A_{*}^{-}}$, i.e.,
\begin{equation}\label{reducer2}
d_{A_{*}} = \lambda [\textbf{1}_{A_{*}^{+}}^t, -\textbf{1}_{A_{*}^{-}}^t]^{t}.
\end{equation}
Upon relabeling, (\ref{K1}) can be equivalently written as
\begin{equation}\label{equv1}
\begin{bmatrix}
\Psi_{A_{*}}^{t} \Psi_{A_{*}} & \Psi_{A_{*}}^{t} \Psi_{I_{*}}\\
\Psi_{I_{*}}^{t} \Psi_{A_{*}} & \Psi_{I_{*}}^{t} \Psi_{I_{*}}
\end{bmatrix}
\begin{bmatrix}
x_{A_{*}}^{*}\\
x_{I-*}^{*}
\end{bmatrix}
+
\begin{bmatrix}
d_{A_{*}}^{*}\\
d_{I_{*}}^{*}
\end{bmatrix}=
\begin{bmatrix}
\Psi_{A_{*}}^t y\\
\Psi_{I_{*}}^t y
\end{bmatrix},
\end{equation}
which, in view of the relations (\ref{reducer1}) and (\ref{reducer2}), can be further rewritten as
\begin{eqnarray}
 \Psi_{A_{*}}^{t} \Psi_{A_{*}} x_{A_{*}}^{*} &=& \Psi_{A_{*}}^t y - d_{A_{*}}^{*},  \label{e29} \\
  d_{I_{*}} &=& \Psi_{I_{*}}^t y-\Psi_{I_{*}}^{t} \Psi_{A_{*}} x_{A_{*}}^{*}.  \label{e210}
\end{eqnarray}

Hence, if the active set $A_{*}$ is known, then the optimal solution $(x^*,d^*)$
follows directly from (\ref{reducer1}), (\ref{reducer2}), (\ref{e29}) and (\ref{e210}).
This motivates a PDAS algorithm. Suppose $x^{k} \in \Rp$ and $d^{k}
\in \Rp$ are approximations to $x^{*}$ and $d^{*}$.
Similar to (\ref{equ:active}), we define the active and inactive sets by
\begin{equation}\label{equ:activek}
\left.\begin{array}{l}A_{k+1}^{+} = \dkh{i \in S: x^{k}_{i} + d^{k}_{i} > \lambda},  \\
A_{k+1}^{-} = \dkh{i \in S: x^{k}_{i} +  d^{k}_{i} < -\lambda},\\
A_{k+1} = A_{k+1}^{+} \cup A_{k+1}^{-},\quad I_{k+1} = A_{k+1}^{c}.
\end{array}\right\}
\end{equation}
Then hopefully, the active set $A_{k+1}$ and inactive set $I_{k+1}$
are also good approximations of  $A^{*}$ and $I^{*}$, respectively. Now
by repeating the arguments leading to (\ref{reducer1}), (\ref{reducer2}),
(\ref{e29}) and (\ref{e210}), we update $x^{k+1} $ and $d^{k+1}$ by the following systems:
\begin{align}
x_{I_{k+1}}^{k+1} &= \textbf{0}_{I_{k+1}},  \label{e211}\\
 d_{A_{k+1}}^{k+1} &= \lambda[\textbf{1}_{A_{k+1}^{+}}^t, -\textbf{1}_{A_{k+1}^{-}}^t]^t,   \label{e212} \\
\Psi_{A_{k+1}}^{t} \Psi_{A_{k+1}} x_{A_{k+1}}^{k+1} &= \Psi_{A_{k+1}}^t y - d_{A_{k+1}}^{k+1},  \label{e213}\\
d_{I_{k+1}}^{k+1} &= \Psi_{I_{k+1}}^t y - \Psi_{I_{k+1}}^{t} \Psi_{A_{k+1}} x_{A_{k+1}}^{k+1}. \label{e214}
\end{align}
Clearly \eqref{e211}, \eqref{e212} and \eqref{e214} involve only matrix-vector multiplications, and thus
they are computationally efficient. The well-posedness of the system depends on
the solvability of \eqref{e213}, which in turn depends on
the property of the submatrix $\Psi_{A_{k+1}}$. In the compressive sensing problem, the active
set $A^*$ is often small. Then if $A_{k+1}$ is an approximation of $A^*$, it is also small and
$\Psi_{A_{k+1}}$ is likely to be a full-column rank matrix. We will discuss the well-posedness
in subsection E below. Now we summarize the PDAS method in Algorithm \ref{PDAS}.

\medskip
\begin{algorithm}[H]
\caption{PDAS}
\begin{algorithmic}[1]
\STATE Input: initial guess $(x^{0},d^{0})$, $\lambda$ and $J$. 
\FOR{$k= 0,1,2,3,\cdots$}
\STATE Compute ${A}_{k+1}$ and $ I_{k+1}$ by \eqref{equ:activek}.
\STATE $ x_{I_{k+1}}^{k+1} = \textbf{0}_{I_{k+1}}.$
\STATE $ d_{A_{k+1}}^{k+1} = \lambda[\textbf{1}_{A_{k+1}^{+}}^t, -\textbf{1}_{A_{k+1}^{-}}^t]^t.$
\STATE $  x_{A_{k+1}}^{k+1} = (\Psi_{A_{k+1}}^{t} \Psi_{A_{k+1}})^{-1}(\Psi_{A_{k+1}}^t y - d_{A_{k+1}}^{k+1}).$
\STATE $ d_{I_{k+1}}^{k+1} = \Psi_{I_{k+1}}^t y - \Psi_{I_{k+1}}^{t} \Psi_{A_{k+1}} x_{A_{k+1}}^{k+1}.$
\STATE Check stopping rule (either $A_k^{\pm} = A_{k+1}^{\pm}$ or $k+1\geq J$).
\ENDFOR
\STATE Output approximation $(x^{k+1},d^{k+1})$.
\end{algorithmic}\label{PDAS}
\end{algorithm}

\subsection{Complexity analysis}
First, we consider the number of floating point operations per iteration. Clearly it takes $O(p)$
flops to finish steps 3 - 5 in the PDAS. In step 6, forming the matrix $\Psi_{A_{k+1}}^{t}
\Psi_{A_{k+1}}$ explicitly takes $O(n|A_{k+1}|^2)$ flops (the cost of forming the
right hand side is negligible since $\Psi^t y$ can be precomputed and retrieved efficiently).
The Cholesky factorization costs $ O(|A_{k+1}|^3)$ flops and the back-substitution needs
$O(|A_{k+1}|^2)$ flops. Hence step 6 takes $O(|A_{k+1}|^2\max(n,|A_{k+1}|))$ flops. At step 7,
two matrix-vector products cost at most $O(np)$ flops.
So, the the overall cost of the PDAS  per iteration is  $O(\max(|A_{k+1}|^3,pn,|A_{k+1}|^2n)$.

The next issue is the number of iterations. Since the PDAS is equivalent to the
semi-smooth Newton method \cite{ssn1,ssn2}, a local superlinear convergence is guaranteed.
The numerical experiments in section 3 also indicate that it converges within a few iterations.
So with a good initial guess, the overall cost of the PDAS is also $O(\max(|A_{k+1}|^3,pn,|A_{k+1}|^2n)$.

If the sought-for solution is sufficiently sparse, i.e., $|A_{k+1}|<\min(n,\sqrt{p})$, the cost
of per PDAS iteration is $O(np)$, which is same as that for other  popular gradient based algorithms.
Moreover, even if the solution is not so sparse, the cost of per PDAS iteration is often
 $O(np)$ by applying Cholesky up/down-date \cite{gloub}. 
To be precise, we downdate by removing the columns in $\Psi_{A_k}$ but not in $\Psi_{A_{k+1}}$
at the cost of $ O(|A_{k}\setminus (A_{k}\cap A_{k+1})||A_{k}|^2)$ flops, and
update by appending the columns in  $\Psi_{A_{k+1}}$ but not in $\Psi_{A_k}$
in $ O(|A_{k+1}\setminus (A_{k}\cap A_{k+1})|(|A_{k}|^2+n|A_{k}|))$ flops.
Then the cost of Cholesky factorization of
$\Psi_{A_{k+1}}^{t}
\Psi_{A_{k+1}}$ is $O((|A_{k}\cup A_{k+1}|-
|A_{k}\cap A_{k+1}|)|A_{k}|(n+|A_{k}|))$. Further, with warm starting, the difference between $A_{k}$
and $A_{k+1}$ is small. Hence, $(|A_{k}\cup A_{k+1}|-|A_{k}\cap A_{k+1}|)$ is not large, and
$(|A_{k}\cup A_{k+1}|-|A_{k}\cap A_{k+1}|)|A_{k}|(n+|A_{k}|) < np$ usually holds.

\begin{remark}
Algorithm 1 requires the explicit form of $\Psi$. Often the signals are sparse
or compressible only in a certain basis. Then the sensing matrix $\Psi$ is the
product of a (random) sampling matrix and the transform matrix, i.e., $\Psi$ is
only given implicitly. One can avoid the explicit expression of $\Psi$ by solving
the linear system at step 6 iteratively, e.g., with conjugate gradient method (CG).
It involves only matrix-vector multiplications, which can often been carried out efficiently
for structured $\Psi$. Only a few CG iterations are needed due to the well-conditionedness of the system.
\end{remark}


%

\subsection{Continuation technique}
In view of the equivalence of the PDAS and the semismooth Newton method\cite{ssn1}, a good initial guess is
essential to its success. For nonsmooth optimization problems, there are several ways to globalize
the Newton method, including squared smoothing with line search \cite{SunQi2001} or path-following
with model function detection \cite{HintermullerKunisch2006}. Due to the special structure of CS problems,
we adopt a continuation technique. Specifically, we consider a decreasing sequence of parameter
$\{\lambda_s\}_s$, and apply Algorithm \ref{PDAS} to $\lambda_{s+1}$-problem with the initial guess from
the solution of $\lambda_{s}$-problem. Summarizing the idea leads to Algorithm \ref{alg:pdasc}.

\begin{algorithm}[H]
\caption{PDASC}\label{alg:pdasc}
\begin{algorithmic}[1]
\STATE Input: $\lambda_0\geq\|\Psi^t y\|_{\infty}$, $ x(\lambda_0) = {\bf 0}_{S}, d(\lambda_0) = \Psi^t y$, $\rho\in (0,1)$.
\FOR{$s= 1,2,3,\cdots$}
\STATE set $\lambda_s = \lambda_0\rho^s$ and $(x^0,d^0) = (x(\lambda_{s-1}), d(\lambda_{s-1}))$.
\STATE Find $x(\lambda_s)$ and $d(\lambda_s)$ by Algorithm 1.
\STATE Check stopping rule.
\ENDFOR
\STATE Output: approximation $(x(\lambda_s),d(\lambda_s))$.
\end{algorithmic}\label{PDASC}
\end{algorithm}

\subsection{ Convergence analysis}
We first consider the convergence of Algorithm \ref{PDAS}. 
The local superlinear convergence of the PDAS can be obtained by reformulating it in the semismooth Newton
framework \cite{ssn1,ssn2,ssn3}. For problems in the CS setting, we can show a stronger result: locally one
step convergence.

\begin{theorem}\label{thm:con}
Let $(x^*,d^*)$ be a solution to the KKT system (\ref{K1})-(\ref{K2}). Suppose the set $\tilde{A}^* =
\{i: |x_i^* + d_i^*| \geq \lambda \}$ is not large in the sense that $\Psi_{\tilde{A}^*}$ is of full
column rank, and the initial guess $(x^0,d^0)$ is close enough to $(x^*,d^*)$. Then $(x^1,d^1)$
generated by Algorithm \ref{PDAS} is $(x^*,d^*)$.
\end{theorem}
\begin{IEEEproof}
See Appendix B.
\end{IEEEproof}

\begin{remark}
The assumption on $\tilde{A}^*$ is closely related to the source condition for the $\ell_1$-minimization problem \cite{Grasmair2011}.
\end{remark}

 Now we show the global convergence of Algorithm \ref{PDASC}. Let $x^\dag$ be the true signal with a support
$A^\dag$ (active set) and the measurement $y$ be noise free, i.e., $y = \Psi x^\dag =
\Psi_{A^\dag} x^\dag_{A^\dag}$. The length of the active set $A^\dag$ is denoted by $T$. The matrix $\Psi$ satisfies
the restricted isometry property (RIP) \cite{cs1}  of order $k$ with constant $\delta_k$  if  $\delta_k \in (0, 1)$  is the smallest constant such that
\begin{equation*}
(1-\delta_k)\|x\|^2 \leq \|\Psi x\|^2 \leq (1+\delta_k)\|x\|^2
\end{equation*}
holds for all $x$ with $\|x\|_0 \leq k$.\\
\noindent \textbf{Assumption 1:}
$\Psi$ satisfies RIP of order $T+1$, and the RIP constant $\delta \triangleq \delta_{T+1} \leq\frac{1}{4\sqrt{T} +1}$.
\begin{theorem}\label{thm:con1}
Let Assumption 1 be fulfilled. With the choice $\rho = \frac{2}{3}$ in Algorithm \ref{PDASC}, and $J\geq T$ in
Algorithm \ref{PDAS}, Algorithm \ref{PDASC} is well-defined. Further, for sufficiently large $s$, the
support of $x(\lambda_s)$ is $A^\dag$, and $\lim_{s\rightarrow \infty} x(\lambda_s) = x^\dag$.
\end{theorem}
\begin{IEEEproof}
See Appendix C.
\end{IEEEproof}

\begin{remark}
\begin{enumerate}
\item Theorem \ref{thm:con1} considers only the noise-free case. In the noisy case,
if the noise level is small, the algorithm is still well defined when equipped with a suitable stopping rule.
\item $J$ is to ensure that Algorithm \ref{PDAS} stops with a finite iteration. In practice, it
is not necessary to be large. 
\item Assumption 1 (with slightly different constant) has been used in the proof of the convergence for orthogonal matching pursuit algorithm (OMP) \cite{Davenport2010}.
\end{enumerate}

\end{remark}
\subsection{Selection of regularization parameter $\lambda$ }
Now we discuss the stopping rule at line 5 of Algorithm \ref{PDASC}
and the choice of regularization parameter $\lambda$.

If the noise level $\epsilon$ is known, the discrepancy principle ($\|\Psi x -y\| \leq \epsilon$) is widely applied to choose a suitable
regularization parameter in inverse problem \cite{Engl1996}. However, for CS problems,
it tends to choose a solution with a very large active set; see the numerical examples in
Section 3. This is attributed to the fact that the $\ell_1$-regularized model may lead a
biased solution \cite{Zhang2008}. More precisely, suppose that the true active set $A^{\dag}$ were
found, and thus primal and dual variables satisfies
\begin{equation*}
d_{A^{\dag}} = \Psi_{A^{\dag}}^t (y - \Psi x), \quad |d_{A^{\dag}}| = \lambda {\bf 1}_{A^{\dag}}.
\end{equation*}
This implies that the residual term $\Psi x - y$ may not be small and hence the discrepancy
principle may not satisfied. Meanwhile, let the oracle solution be
$$x^o_{A^{\dag}} \triangleq \Psi_{A^{\dag}}^\dag y = (\Psi_{A^{\dag}}^t\Psi_{A^{\dag}})^{-1}\Psi_{A^{\dag}}^t y.$$
Then on the active set, there holds $x_{A^{\dag}} + (\Psi_{A^{\dag}}^t\Psi_{A^{\dag}})^{-1}d_{A^{\dag}} = x^o_{A^{\dag}}$.
Hence, $x_{A^{\dag}} + (\Psi_{A^{\dag}}^t\Psi_{A^{\dag}})^{-1}d_{A^{\dag}} $ is a better approximation to the
true solution. This motivates us to propose a modified discrepancy principle (MDP)
for the stopping rule and selecting the regularization parameter. Specifically, let the
active set of $x(\lambda_s)$ in Algorithm \ref{PDASC} be $A_{s}$. Algorithm \ref{PDASC} stops when
\begin{equation*}
 \|\Psi_{A_{s}} (x(\lambda_s)_{A_{s}} + (\Psi_{A_{s}}^t\Psi_{A_{s}}
 )^{-1}d(\lambda_s)_{A_{s}}) - y\|\leq \epsilon,
\end{equation*}
where $\epsilon$ is the noise level, and accordingly the approximate solution is given by
$$x_{A_{s}} = x(\lambda_s)_{A_{s}} + (\Psi_{A_{s}}^t\Psi_{A_{s}})^{-1}d(\lambda_s)_{A_{s}},\quad \mbox{and}\quad
 x_{I_{s}} = {\bf 0}_{I_{s}}.$$
The above equation is a debias step, see also \cite{gpsr} for the similar debias postprocessing.
 One should be noticed that after this debias postprocess, the  solution obtained may  not be the solution to \eqref{rgl1*}, but be more closed to solution of $\ell_0$-minimization problem.  This debias postprocess will only  been done when the modified discrepancy principle (MDP) is satisfied.

If the noise level is unknown, we choose the stopping criterion at line $5$ of Algorithm \ref{PDASC}
as the size of the active set, e.g., $\|x(\lambda_s)\|_0 \geq \eta n$ for $\eta \in
[0.5,1]$. To choose a proper regularization parameter $\lambda$, we employ Bayesian
information criterion (BIC), which is a data driven method and  widely used in statistics
due to its  model selection consistency \cite{bic2,bic5}. BIC chooses $\lambda$ by:
   \begin{equation}\label{bic}
   \min_{\lambda\in \Lambda} \left\{BIC(\lambda):= \frac{1}{2}\|\Psi x_{\lambda}-y\|_{2}^2  + \frac{\ln n}{n}df_{\lambda}\right\},
   \end{equation}
where $x_{\lambda}$ is the solution of (\ref{rgl1*}), $\Lambda$ is a subset of $(0,+\infty)$,
and $df_{\lambda}$ represents the degree of freedom of $x_{\lambda}$ that can be chosen as
$\|x_{\lambda}\|_0$ \cite{df}. Due to the complex structure of the BIC functional, it is
nontrivial to find its minimizer over the whole positive real line. Instead, a practical way
is to find the minimizer over the finite candidate set $\Lambda = \{\lambda_s\}_s$ which will be specified in the next section in numerical tests.

\section{Numerical Examples}

Now we present numerical examples to show the efficiency and accuracy of Algorithm \ref{alg:pdasc} (PDASC).
First, we give the implementation details, e.g., the generation of
simulation data, parameter setting for the algorithm. Then we check the efficiency of regularization parameter choice strategy: for both MDP and BIC based parameter choice rules. Later on our method is also compared
with several state-of-the-art algorithms for both CPU time and reconstruction error.

\subsection{Implementation Setting}
The signals $x^\dag$ are chosen as $T$-sparse with a dynamic range
$$Dyna := \frac{\max\{|x^\dag_{i}|:x^\dag_{i}\neq 0\}}{\min\{|x^\dag_{i}|:x^\dag_{i}\neq 0\}}.$$
They are generated following \cite{nesta}.

The sensing matrix $\Psi$ of size $n\times p$ is chosen to be either random Gaussian matrix,
or random Bernoulli matrix, or partial discrete cosine transform (DCT) matrix. The observation vector $y$ is given by
$y = \Psi x^\dag+\eta$, where  $\eta$ is the  Gaussian noise vector whose entries are i.i.d. $\sim N(0,\sigma)$.

One needs the following algorithm parameters: initial regularization parameter $\lambda_0$; decreasing factor $\rho$; maximal iteration number $J$, and noise level $\epsilon$. The noise level is chosen as $\epsilon = \|\eta\|$. The maximal iteration number $J$ is not sensitive to the algorithm (due to the locally superlinear or one step  convergence property of PDAS), one can choose it as $J=1$. To determine the initial regularization parameter $\lambda_0$ and decreasing factor $\rho$, we pickup an interval $[\lambda_{min},\lambda_{max}]$ which contains the target  the regularization parameter. Then an equal-distributed partition on log-scale is employed to divide this interval into $N$-subintervals. Clearly larger $N$ implies larger $\rho$. For simplicity, let $\lambda_{max} = \|\Psi^t y\|_{\infty}$, $\lambda_{min} = \textrm{1e-10} \lambda_{max}$ and $N=100$.

When the sensing matrix $\Psi$ is random Gaussian matrix or random Bernoulli matrix, the matrix $\Psi^t\Psi$ is saved in advance (not be included in CPU time), and the linear equation in line 6 of Algorithm \ref{PDAS} is solved by Cholesky factorization. But when $\Psi$ is a partial discrete cosine transform matrix, we do not have the explicit form of $\Psi$ and $\Psi^t$.   The linear equation in line 6 of Algorithm \ref{PDAS} is solved by conjugated gradient (CG) method initialized with the projection of the previous solution onto the current active set.  We set the number of  CG iteration as 2 in all the simulations below.

\subsection{Check regularization  parameter selection rules}
We will check the ability of proposed regularization selection rules. The three rules are compared in Table \ref{tab:1}, they are modified discrepancy principle (MDP), Bayesian information criterion (BIC), and standard discrepancy principle (DP). We consider nine different cases where the sensing matrix $\Psi$ is chosen as $512\times 2048$ partial DCT matrix, $256\times 1024$ random Gaussian matrix, and $200\times 1000$ random Bernoulli matrix, respectively. For each type of sensing matrix $\Psi$, we consider three different noise level and different sparsity level. The details is shown  in Table \ref{tab:1}.

The first two columns of Table \ref{tab:1} are different problem setting and different parameter selection rules. The third and forth columns are the CPU time (in seconds) and relatively $\ell_2$ error. The fifth and sixth columns are information of active set. Column five and six are the size of $\hat{A}\setminus A^{\dag}$ and $A^{\dag}\setminus \hat{A}$, respectively, where $\hat{A}$ and $A^\dag$ be the numerical active set and true active set. The last column is the selected regularization parameter $\hat{\lambda}$. In some cases DP may fail and we use $F$ to indicate it.

It should be noticed that the regularization parameter from MDP is much larger than the ones from BIC or DP. The reason is that   Algorithm 2 stops immediately when active set $\hat{A}$ contains  the true active set $A^\dag$  due to the debias step. Hence the debias postprocess makes  Algorithm 2  terminate earlier and  make larger $\lambda$ to be selected.  One can find in Table \ref{tab:1} that when noise level and sparsity level are small, three methods all work well. When the noise level and sparsity level are relatively large, DP may fail, MDP and BIC still work. In most cases, MDP takes less CPU time and chooses a smaller (more accurate) active set, but it requires the information of noise level. In later numerical tests, if the noise level is known, we can use either MDP and BIC to find a solution, otherwise only BIC is available.

\begin{table}[h]
\caption{ Comparison for choosing regularization parameter}\label{tab:1}
\begin{center}
\begin{tabular}{ccccp{0.3cm}p{0.3cm}c}
\hline\hline
\multicolumn{1}{c}{setting} & \multicolumn{1}{c}{method} & \multicolumn{1}{c}{time(s)} & \multicolumn{1}{c}{error} & \multicolumn{2}{c}{active set} & \multicolumn{1}{c}{$\hat{\lambda}$}
\\
 \hline
    Partial DCT      &MDP    &0.20     &4.66e-5        &0       &0       &1.80e-1   \\
    $\sigma =$ 1e-4  &BIC    &0.49     &3.83e-4        &0       &0       &3.79e-4   \\
   $|A^{\dag}|=32$   &DP     &0.39     &1.37e-4         &106     &0       &1.05e-4 \\
 \hline
  Partial DCT       &MDP   &0.23     & 4.57e-3        &9       &0       &1.62e-1           \\
  $\sigma =$ 1e-2   &BIC   &0.28     & 3.96e-2        &13       &0       &3.57e-2      \\
 $|A^{\dag}|=64$    &DP    &0.26     &1.50e-2         &173     &0       &1.13e-1    \\
 \hline
  Partial DCT      &MDP   &0.22     &5.48e-2         &50       &2       &1.79e-1 \\
  $\sigma =$ 5e-2  &BIC    &0.19     &9.89e-2         &85       &0        &7.92e-2        \\
 $|A^{\dag}|= 80$   &DP   &F         &F                &F       &F        &F    \\

 \hline
  Gaussian         &MDP    &3.2e-2     &2.10e-5      &0       &0       &8.72e-1     \\
 $\sigma =$ 1e-4   &BIC    &3.5e-2     &1.23e-4        &0       &0       &5.87e-3   \\
  $|A^{\dag}|= 16$  &DP     &4.9e-2     &5.24e-5        &47      &0        &2.01e-4 \\
 \hline
 Gaussian           &MDP   &1.1e-2    & 2.34e-3      &10       &0       &5.92e-1           \\
 $\sigma =$ 1e-2      &BIC   &2.6e-2    & 1.56e-2        &14       &0       &6.50e-2      \\
 $|A^{\dag}|= 32$   &DP    &2.4e-2     &6.02e-3              &87         &0       &1.61e-2\\
 \hline
 Gaussian            &MDP   &1.4e-2     &2.00e-2      &31       &0       &4.62e-1 \\
 $\sigma =$ 5e-2       &BIC   &2.1e-2    &5.90e-2        &54       &0       &1.62e-1    \\
 $|A^{\dag}|= 40$     &DP      &F       &F             &F        &F       &F     \\
 \hline
   Bernoulli         &MDP    &1.2e-2     &1.96e-6      &0       &0       &8.53e-1     \\
   $\sigma =$ 1e-3   &BIC    &2.4e-2    &2.19e-5        &0       &0       &1.38e-3   \\
   $|A^{\dag}|= 10$  &DP      &2.1e-2    &4.82e-5       &41    &0      &2.02e-3\\
 \hline
  Bernoulli            &MDP   &1.5e-2     & 2.96e-4        &8       &0       &6.91e-1           \\
  $\sigma =$ 1e-2        &BIC   &2.4e-2     & 2.22e-3        &9       &0      &8.51e-2      \\
  $|A^{\dag}|= 25$       &DP    &2.3e-2        &6.70e-4           &78    &0      &1.68e-2\\
  \hline
  Bernoulli          &MDP   &1.8e-2     &1.10e-2         &48       &0        &6.26e-1 \\
  $\sigma =$ 1e-1     &BIC   &2.0e-2     &2.10e-2         &71       &0        &3.93e-1    \\
  $|A^{\dag}|= 40$  &DP    &F    &F       &F    &F   &F   \\

\hline\hline
\end{tabular}
\end{center}
\end{table}

\subsection{Comparison with other algorithms}

We compare our algorithm with the several state-of-the-art algorithms
for solving \eqref{rgl1*}. The parameters in these algorithms are
the default values as their online packages,
except for the stopping criterion which will be discussed later.

Gradient projections for sparse reconstruction (GPSR) \cite{gpsr} uses Barzilai-Borwein
rule to choose step length. The MATLAB code is available at http://www.lx.it.pt/mtf/GPSR/.

The Matlab code for sparse reconstruction by separable approximation (SpaRSA) \cite{sparsa}
is available at http://www.lx.it.pt/mtf/SpaRSA/.

The package of fixed point continuation (FPC) \cite{fpc} and its modified version (FPC-AS) \cite{wen2010fast} are available at http://www.caam.rice.edu/$\sim$optimization/L1/.

For all these algorithms, a regularization parameter is needed. Since the solution by MDP is slightly different from the solution to (\ref{rgl1*}), we use BIC to pickup a regularization parameter and use it in other algorithms.

\begin{table}[h]
  \caption{Random Bernoulli matrix}\label{tab:2}
  \vspace{-0.5cm}
  \begin{center}
  \begin{tabular}{ccccccc}
  \hline
     method      &Time & $\ell_2$ RE &  $\ell_\infty$ AE & $\ell_2$ dRE &  $\ell_\infty$ dAE    \\
 \hline
   PDASC-l1(MDP) &1.85  &4.80e-6 &3.10e-3           &4.80e-6 &3.10e-3             \\
   PDASC-l1(BIC) & 3.58 &8.90e-5 &3.96e-2           &5.14e-6 &3.28e-3           \\
   GPSR-bb       & 9.82 &1.45e-4 &7.18e-2           &7.07e-6 &3.65e-3              \\
   SpaRSA        & 10.1 &1.05e-4 &5.12e-2           &5.64e-6 &3.44e-3           \\
   FPC           &42.9  &2.69e-4 &1.27e-1           &2.54e-4 &1.15e-1              \\
   FPC-AS        &5.57  &9.56e-5 &4.32e-2           &3.83e-6 &2.81e-3                  \\
   \hline
  \end{tabular}
  \end{center}
 \center {$n=2048$, $p=32768$, $T=128$, $\text{Dyna}$=1e3, $\sigma = 1e-3$.}
\end{table}

As was pointed out in \cite{nesta}, to compare different algorithms, one needs a
fair stopping criterion. We setup the stop condition for other algorithm as follows.
Firstly we use BIC to get a regularization parameter $\hat{\lambda}$ and a solution $x({\hat{\lambda}})$.
Then the stopping rule for other $\ell_{1}$ solvers is either their default stop criterions or the following condition is fulfilled:
\begin{equation*}
  \tfrac{1}{2}\|\Psi x^{k}-y\|^{2}_{2} + \hat{\lambda}\|x^{k}\|_{1}\leq \tfrac{1}{2}\|\Psi x({\hat{\lambda}})-y\|^{2}_{2} + \hat{\lambda}\|x({\hat{\lambda}})\|_{1}.
\end{equation*}

The first group experiments are to recover three different $T$-sparse signal
with $T = 128,256,1024$, which are sampled by random Bernoulli  matrix with
size $4096\times16384$, random Gaussian  matrix with size $2048\times 32768$,
and partial DCT matrix with size $16384\times65536$, respectively. The dynamic range
of in those tests are 1e3, 1e4, 1e2, respectively. The noise $\sigma$ is chosen as 1e-3, 1e-2, 1e-2, respectively.
The  averaged   results  based on of $10$ independent replications
(CPU times,  $\ell_{2}$ relative errors ($\ell_{2}$ RE), $\ell_{\infty}$
absolute errors ($\ell_{\infty}$ AE), $\ell_{2}$ relative errors  after debias ($\ell_{2}$ dRE) and  $\ell_{\infty}$
absolute errors after debias ($\ell_{\infty}$ dAE) ) are reported in Tables \ref{tab:2} - \ref{tab:4}.

\begin{table}[h]
  \caption{Random Gaussian matrix}\label{tab:3}
  \vspace{-0.5cm}
  \begin{center}
  \begin{tabular}{ccccccc}
  \hline
     method      &Time & $\ell_2$ RE &  $\ell_\infty$ AE & $\ell_2$ dRE &  $\ell_{\infty}$ dAE\\
 \hline
   PDASC-l1(MDP) &3.02 &4.66e-6       &3.32e-2  &4.66e-6       &3.32e-2     \\
   PDASC-l1(BIC) &4.53 & 1.53e-5   &6.23e-2  &1.26e-5       &5.47e-2    \\
   GPSR-bb       &6.37 & 1.83e-5   &8.74e-2  &1.81e-5       &6.95e-2    \\
   SpaRSA        &10.2 & 1.55e-5   &6.62e-2  &1.35e-5       &5.49e-2    \\
   FPC           &25.4 & 3.52e-5   &9.24e-2  &1.96e-5       &9.17e-2    \\
   FPC-AS        &6.19 & 1.83e-5   &7.74e-2  &1.59e-5       &6.86e-2    \\
   \hline
  \end{tabular}
  \end{center}
 \center {$n=4096$, $p=16384$, $T=256$, $\text{Dyna}$=1e4, $\sigma=1e-2$.}
\end{table}

\begin{table}[h]
  \caption{partial DCT  matrix}\label{tab:4}
  \vspace{-0.5cm}
  \begin{center}
  \begin{tabular}{ccccccc}
  \hline
     method      &Time & $\ell_2$ RE &  $\ell_{\infty}$ AE & $\ell_2$ dRE &  $\ell_{\infty}$ dAE \\
 \hline
 PDASC-l1(MDP)   & 1.56  &6.54e-4 &0.08              &6.54e-4 &0.08              \\
 PDASC-l1(BIC)   & 1.02  &2.04e-3&0.13              &1.94e-3&0.11              \\
   GPSR-bb       & 0.87  &2.10e-3&0.14              &2.04e-3&0.11              \\
   SpaRSA        & 1.14  &2.01e-3&0.13              &1.95e-3&0.11              \\
   FPC           & 0.76  &2.17e-3&0.15              &2.19e-3&0.12             \\
   FPC-AS        & 0.68  &2.05e-3&0.12              &1.60e-3&0.10             \\
   \hline
  \end{tabular}
  \end{center}
 \center {$n= 16384$, $p=65536$,$T= 1024$,  $\text{Dyna}$=1e2, $\sigma=1e-2$.}
\end{table}

In Table \ref{tab:2} - \ref{tab:4}, PDASC with MDP (the noise level is supposed to known) or BIC are compared with other four algorithms. The first two columns are method and CPU time (in seconds), and last four columns are errors of the solutions.  Columns three and four are standard relatively $\ell_2$ error and absolute $\ell_\infty$ error. The last two columns are the $\ell_2$ and $\ell_\infty$ after a debias postprocess. It is observed that Algorithm \ref{alg:pdasc} is very competitive to  other state-of-art algorithms in both accuracy and CPU time. However,  the regularization parameter is not necessarily known in advance for PDASC which may  make PDASC a good candidate for  for large scale real data. If the sensing matrix is random Bernoulli or random Gaussian, PDASC with MDP is fastest, and  when $\Psi$ is partial DCT matrix PDASC with MDP is a bit slower. This fact is due to that we apply different solvers for the linear system in step 6 of Algorithm 1, i.e., Cholesky factorization for previous two cases (the explicit form of $\Psi^t\Psi$ is needed) and CG for the last case, respectively.

\begin{table}[h]
 \caption{ One dimensional signal}\label{tab:5}
  \begin{center}
 \begin{tabular}{ccccc}
 \hline
    method   &CPU time  &PSNR      \\
 \hline
  PDASC-l1      & 0.50  &54     \\
  GPSR-bb  & 0.62  &54     \\
  SpaRSA    & 0.70  &54       \\
  FPC        & 0.42  &54    \\
   FPC-AS        & 0.70 &54     \\
  \hline
  \end{tabular}
 \end{center}
 \center {$n=665$, $p=1024$, $T=247$,  $\sigma$=1e-4, $\hat{\lambda}$=7.42e-4.}
\end{table}

\begin{figure}[h]
\centering
\includegraphics[trim = 1.8cm 2cm 1.2cm 0cm, clip=true,width=8cm]{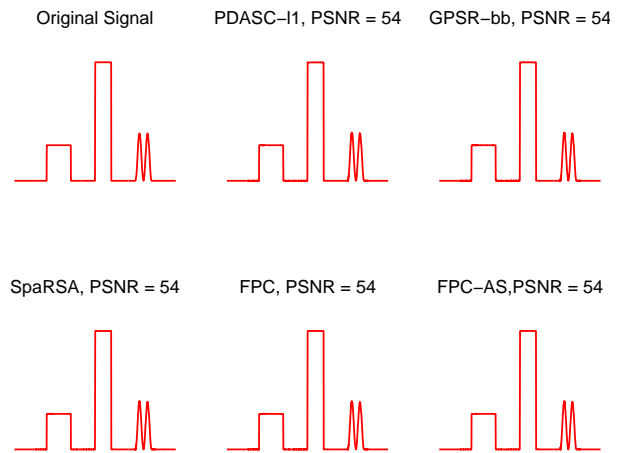}
\caption{Reconstruction signal and their PSNR values  }\label{fig:1}
\end{figure}

\begin{figure}[h]
\centering
\includegraphics[trim = 1.8cm 1cm 1.1cm 0cm, clip=true,width=8cm]{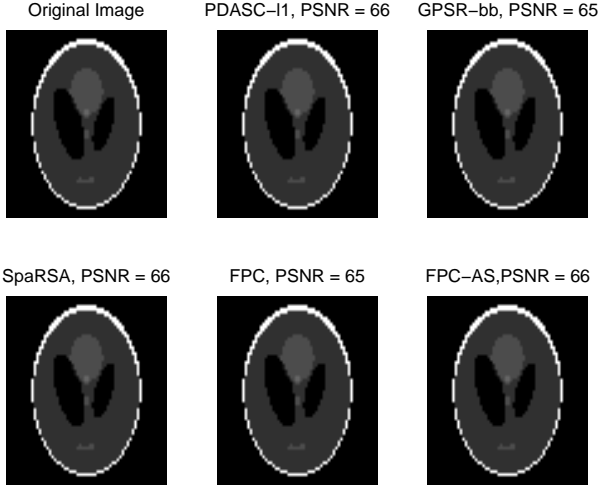}
\caption{Reconstruction Phantom images and their PSNR values  }\label{fig:2}
\end{figure}

Next group of numerical examples  reconstruct a one dimensional signal and a benchmark MRI image. Both of them are compressible under
a Haar  wavelet basis. Therefor, the observation data can be chosen as the wavelet coefficients sampled by the product of a partial FFT matrix and inverse Haar wavelet transform.  Similarly one
needs a regularization parameter for other state-of-the-art algorithms.
Same as before, we first run Algorithm \ref{alg:pdasc} with BIC to get a
regularization parameter $\hat{\lambda}$, and use it for  other $\ell_{1}$ solvers.
 In these two examples  we assume the noise level is not known (this is the case for most real data) and  we use PDASC with BIC to compare with other solver by CPU time and PSNR value. The stopping rule for other algorithms are the same
as before. The results are reported in Table \ref{tab:5}, \ref{tab:6} and Figure \ref{fig:1}, \ref{fig:2}.

\begin{table}[h]
 \caption{Two dimensional imagine }\label{tab:6}
  \begin{center}
 \begin{tabular}{ccccc}
 \hline
    method   &CPU time  &PSNR      \\
 \hline
  PDASC-l1      & 0.52  &66     \\
  GPSR-bb  & 0.76  &65     \\
   SpaRSA    & 0.86  &66       \\
  FPC        & 0.92  &65    \\
   FPC-AS        & 1.75 &66     \\
  \hline
  \end{tabular}
 \end{center}
 \center {$n=2133$, $p=4096$, $T=792$,  $\sigma$=1e-4, $\hat{\lambda}$=5.35e-4.}
\end{table}

For the one dimensional signal, the sampling matrix $\Psi$ with  size $665\times 1024$ is the compound of a partial FFT and  a inverse wavelet transform, and the signal under wavelet transformation has $247$ nonzero entries. The sampling matrix $\Psi$ for two dimensional MRI imagine is the compound of a partial FFT and an inverse wavelet transform with size $2133\times 4096$. The image under wavelet transformation has $792$ nonzero entries. The numerical results also demonstrate that    the proposed PDASC is very competitive in terms of efficiency and accuracy, but without a priori knowledge of regularization parameter.

\section{Conclusion}
A primal dual active set with continuation algorithm together with suitable regularization
parameter choice rules has been proposed to solve $\ell_1$-regularized least squares problem.
We derived the local one step convergence of PDAS and established the global convergence of PDASC.
Numerical experiments verified the algorithm PDASC is very competitive to the state-of-art $\ell_1$ solvers both in  accuracy  and  efficiency.
 There are several questions deserving further study. First, if the sensing matrix is
implicit given (such as partial DCT matrix) an iterative solver is needed in each Newton step. A proper stopping rule for this
inner iteration is important and remains unclear. Second, BIC is very promising  data driven
regularization parameters selection rule, but its efficient implementation is still challenging. Last,
 adaptation of the algorithm to more complicated scenarios, such as severely ill-posed inverse
problems, is also of immense practical interest.

\appendices
\section{Proof of Theorem 1}

\begin{IEEEproof}
Let $x^{*} \in \Rp$ be a minimizer of (\ref{rgl1*}), then by (\ref{fermat}) we have
\begin{equation}\label{eqapp}
\textbf{0}  \in \Psi^t(\Psi x^{*} - y) + \lambda\partial\|\cdot\|_{1}(x^{*}).
\end{equation}
Therefore, there exists $d^{*} \in \lambda\partial\|\cdot\|_{1}(x^{*})$ such that $\textbf{0}  = \Psi^t(\Psi x^{*} - y) + d^{*}$. Now (\ref{general}) and (\ref{proxl1}) imply,
$$d^{*} \in \lambda\partial\|\cdot\|_{1}(x^{*}) \Leftrightarrow x^{*} = Prox_{\lambda\|\cdot\|_{1}}(x^{*}+ d^{*})  = T_{\lambda}(x^{*} + d^{*}).$$

Conversely, suppose (\ref{K1}) and (\ref{K2}) hold.
By (\ref{general}) and (\ref{K1}), we obtain that $d^{*} \in\lambda\|\cdot\|_{1}(x^{*})$.
Substitute to (\ref{K1}) we have $\textbf{0} \in \Psi^t(\Psi x^{*} - y) + \lambda \|\cdot\|_{1}(x^{*})$.
By Fermat's rule (\ref{fermat}), we conclude that $x^{*}$ is a minimizer of (\ref{rgl1*}).
\end{IEEEproof}

\section{Proof of Theorem 2}
\begin{IEEEproof}
Let $J^* = \{i: |x_i^* + d_i^*| \neq \lambda\}$, and
\begin{equation*}
  \theta = \min_{i\in J^*} ||x_i^* + d^*_i| - \lambda| > 0.
\end{equation*}
We assume that the initial guess $(x^0,d^0)$ is close to $(x^*,d^*)$ in the sense
$\|x^* - x^0\|_{\infty} + \|d^* - d^0\|_{\infty} \leq \theta.$

Like $A^{\pm}_*$ and $A^{\pm}_1$ in (\ref{equ:active}) and (\ref{equ:activek}), we denote by
$\tilde{A}^{+}_* = \{i: x_i^* + d_i^* \geq \lambda\}$, $\tilde{A}^{-}_* = \{i: x_i^* + d_i^* \leq -\lambda\},$
and $\tilde{A}^* = \tilde{A}^{+}_* \cup \tilde{A}^{-}_*$.
For any $i=1,...,p$, there holds
$$|(x^0_i + d^0_i) - (x^*_i + d^*_i)| \leq \|x^* - x^0\|_{\infty} + \|d^* - d^0\|_{\infty} \leq \theta.$$
This relation with definition of $\theta$ implies that
\begin{eqnarray*}
&&x_i^* + d_i^* \gtrless \pm \lambda \Rightarrow x_i^0 + d_i^0 \gtrless \pm \lambda,\\
&&|x_i^* + d^*_i| < \lambda \Rightarrow |x_i^0 + d^0_i| < \lambda,
\end{eqnarray*}
and hence
 $A^{\pm}_* \subseteq A^{\pm}_1\subseteq \tilde{A}^{\pm}_*$. From the definition
of $\tilde{A}^{\pm}_*$, we notice that $d^*|_{\tilde{A}^{\pm}_*} = \pm \lambda$. Combining it with (\ref{e212}) yields
\begin{equation*}
d^*|_{\tilde{A}^{\pm}_*} = \pm \lambda \Rightarrow d^*|_{A^{\pm}_1} = \pm \lambda = d^1|_{A^{\pm}_1}.
\end{equation*}
Using (\ref{K1}), (\ref{e213}), and the relation $\Psi x^* = \Psi_{A_1} x^*_{A_1}$, we deduce
\begin{equation*}
\Psi_{A_1}^t\Psi_{A_1} x^*_{A_1} + d^*_{A_1} = \Psi^t_{A_1} y = \Psi_{A_1}^t\Psi_{A_1} x^1_{A_1} + d^1_{A_1},
\end{equation*}
which implies that $\Psi_{A_1}^t\Psi_{A_1}( x^*_{A_1} -  x^1_{A_1}) = 0$. Since $\Psi_{\tilde{A}^*}$
has a full column rank, $\Psi_{A_1}^t\Psi_{A_1}$ is  invertible and thus $ x^*_{A_1} = x^1_{A_1} $.
By $x^*_{I_1} = {\bf 0}_{I_1} = x^1_{I_1}$, we conclude the desired result $x^1 = x^*$.
\end{IEEEproof}

\section{Proof of Theorem \ref{thm:con1}}

We first recall some standard estimates for RIP constants \cite{cosamp}.
Let $A,B$ be disjoint subsets of $\{1,2,...,p\}$, then
\begin{eqnarray*}
&&\|\Psi_A^t\Psi_A x_A\|\gtreqqless (1\mp \delta_{|A|})\|x_A\|,\\
&&\|(\Psi_A^t\Psi_A)^{-1} x_A\|\gtreqqless \frac{1}{1\mp \delta_{|A|}}\|x_A\|,\\
&&\|\Psi_A^t\Psi_B\|\leq \delta_{|A|+|B|},\\
&&\|\Psi_A^+ y\|\leq \frac{1}{\sqrt{1-\delta_{|A|}}}\|y\|.
\end{eqnarray*}

Now we give a few more preliminary estimates. Let $A \subset A^\dag$,
$ I = A^c$ and $ B = A^\dag\backslash A$, and consider one step iteration:
\begin{equation*}
\left\{\begin{array}{l}
x_{ I} = \mathbf{0}_{ I}, \quad |d_{A}| = \lambda \mathbf{1}_{A}, \\[1.5ex]
x_{A} = (\Psi_{A}^t\Psi_{A})^{-1}(\Psi_{A}^t y - d_{A}),\\[1.5ex]
d_{ I} = \Psi^t_{ I}(y - \Psi_{A}x_{A}).
\end{array}\right.
\end{equation*}
Upon noting $y = \Psi_{A^\dag}x_{A^\dag}^\dag$ and $A^\dag = A \cup  B$, we deduce
\begin{equation*}
x_{A} = (\Psi^t_A\Psi_{A})^{-1}(\Psi_A^t(\Psi_{A}x_{A}^\dag + \Psi_{ B}x_{ B}^\dag) - d_A),
\end{equation*}
and hence
\begin{equation*}
\begin{aligned}
\|x_{A} + d_{A} - x_{A}^\dag\| &\leq \|(\Psi^t_A\Psi_{A})^{-1}\Psi_A^t\Psi_{ B}x_{ B}^\dag\| \\
&\qquad + \|(I - (\Psi^t_A\Psi_{A})^{-1})d_A\| \\ 
&\leq \tfrac{\delta}{1-\delta}\|x_{ B}^\dag\| + \tfrac{\delta}{1-\delta}\|d_{A}\|.
\end{aligned}
\end{equation*}
In view of the relation
\begin{equation*}
  \begin{aligned}
    d_i &= \Psi^t_i(y  - \Psi_A x_{A}) \\
      &= \Psi^t_i(\Psi_A(x_{A}^\dag - x_A - d_A) + \Psi_Ad_A + \Psi_{B}x_ B^\dag),
  \end{aligned}
\end{equation*}
for any $i\in I^\dag$ we have
\begin{align*}
|d_i| &\leq \delta(\|x_{ B}^\dag\| + \|d_{A}\| + \|x_{A}^\dag - x_A - d_A\|) \\
& \leq \tfrac{\delta}{1-\delta}\|x_{ B}^\dag\| + \tfrac{\delta}{1-\delta}\|d_{A}\|.
\end{align*}
Let $i_A \in \mathop\textrm{arg}\max\limits_{i\in I}|x^\dag_i|$. Clearly $i_A\in B$, and hence
\begin{align*}
  |d_{i_A}|& \geq |x_{i_A}^\dag| - \delta(\|x_{ B}^\dag\| + \|d_{A}\| + \|x_{A}^\dag - x_A - d_A\|) \\
  &\geq |x_{i_A}^\dag| - \tfrac{\delta}{1-\delta}\|x_{ B}^\dag\| - \tfrac{\delta}{1-\delta}\|d_{A}\|.
\end{align*}
By the trivial estimates $\|x_ B\| \leq \sqrt{| B|}|x_{i_A}^\dag|$, $\|d_{A}\| = \sqrt{|A|}\lambda$, and
the implication $\delta\leq\frac{1}{4\sqrt{T}+1} \Rightarrow \frac{\delta\sqrt{T}}{1-\delta}\leq \frac{1}{4}$, we deduce
\begin{align}
  \|x_{A} + d_{A} - x_{A}^\dag\| & \leq \tfrac{1}{4}|x_{i_A}^\dag| + \tfrac{1}{4}\lambda, \label{ine:a1}\\
  |d_{i_A}| &\geq \tfrac{3}{4}|x_{i_A}^\dag| - \tfrac{1}{4}\lambda, \label{ine:a2}\\
  |d_i| &\leq \tfrac{1}{4}|x_{i_A}^\dag| + \tfrac{1}{4}\lambda, \quad \forall i\in I^\dag. \label{ine:a3}
\end{align}
Further, for any given $\lambda>0$ and $m>0$, we define the set
\begin{eqnarray}\label{equ:jm}
J_{\lambda,m} = \{i: |x_i^\dag|\geq m\lambda\}.
\end{eqnarray}

The proof of Theorem \ref{thm:con1} is based on the following claim for one
iteration of Algorithm \ref{PDAS}.\\
\noindent{Claim 1:} Let $m=2$ or $3$. \\
a. If $J_{\lambda,2}\subseteq A_k\subseteq A^\dag$, then $J_{\lambda,2}\subseteq A_{k+1}\subseteq {A}^\dag$. \\
b. If $J_{\lambda,3}\subseteq A_k\subseteq A^\dag$, we have
either $J_{\lambda,2}\subseteq A_{k}$ or
\begin{equation*}
\max\{|x^\dag_i|: i\in I_k\} > \max\{|x_i^\dag|:i\in I_{k+1}\}.
\end{equation*}
\noindent{Proof:} By the assumption $J_{\lambda,m}\subseteq A_k\subseteq A^\dag$, we
have $|x_{i_{A_k}}^\dag| < m\lambda$. Combining estimates (\ref{ine:a1})-(\ref{ine:a3}) yields
for $m=2,3$
\begin{align*}
 &|x^k_i + d^k_i| \geq \tfrac{3}{4}|x_i^\dag| - \tfrac{1}{4}\lambda \geq \tfrac{3m-1}{4}\lambda > \lambda, \forall i\in J_{\lambda,m},\\
 &|d_i| \leq \tfrac{1}{4}|x_{i_{A_k}}^\dag| + \tfrac{1}{4}\lambda < \tfrac{m+1}{4}\lambda < \lambda, \forall i\in I^\dag,
\end{align*}
which implies that $J_{\lambda,m}\subseteq A_{k+1}\subseteq {A}^\dag$. Now we assume $J_{\lambda,3}\subseteq A_k\subseteq A^\dag$
and $J_{\lambda,2} \nsubseteq A_{k}$. Then for any $i_{A_k}$ in $ J_{\lambda,2}\backslash J_{\lambda,3} $,
$|x_{i_{A_k}}^\dag|\in [2\lambda, 3\lambda)$. Consider any $i\in A_k$ such that $|x_{i}^\dag| \geq |x_{i_{A_k}}^\dag|$, we have
\begin{equation*}
|x^k_i + d^k_i| \geq \tfrac{3}{4}|x_i^\dag| - \tfrac{1}{4}\lambda >\lambda \Rightarrow i\in A_{k+1}.
\end{equation*}
For any $i_{A_k}$, we also have
\begin{equation*}
|d_{i_{A_k}}| \geq \tfrac{3}{4}|x_{i_{A_k}}^\dag| - \tfrac{1}{4}\lambda > \lambda\Rightarrow {i_{A_k}}\in A_{k+1}.
\end{equation*}
Therefore $\max\{|x^\dag_i|: i\in I_k\} > \max\{|x_i^\dag|:i\in I_{k+1}\}$.

Now we state the proof of Theorem \ref{thm:con1}.
\begin{IEEEproof}
For any given $\lambda_s$, let Algorithm \ref{PDAS} take $k_s$-steps to stop and denote the active
set during the PDAS iteration (cf. Algorithm \ref{PDAS}) by $A_{k,s}$ for $k\leq k_s$, and
\begin{equation*}
A_{\diamond,s} = \{i:|x_i^{k_s} + d_i^{k_s}| > \lambda_s\}.
\end{equation*}
By construction (cf. Algorithm \ref{PDASC}), we have $(x^{k_s}, d^{k_s}) = (x(\lambda_s),d(\lambda_s))$,
and it is the initial guess for $\lambda_{s+1}$-problem. We shall prove $A_{k,s}\subseteq A^\dag$ by
mathematical induction and hence also the well-posedness of the algorithm. To this end, we need the
following claim:\\
\noindent {Claim 2:} For any $s\geq 0$, we have $J_{\lambda_s,3}\subseteq A_{1,s}\subseteq A^\dag$ and $J_{\lambda_s,2}\subseteq A_{\diamond,s}\subseteq A^\dag$.\\
\noindent \textit{Step 1.} For any $s\geq 0$, if $J_{\lambda_s,3}\subseteq A_{1,s}\subseteq A^\dag$,
then by Claim 1, we have $J_{\lambda_s,3}\subseteq A_{k,s}\subseteq A^\dag$ for any $k\leq k_s$. When
Algorithm \ref{PDAS} stops, it is either $A_{k_s,s}^{\pm} = A_{k_s+1,s}^{\pm}$ or $k_s = J\geq T$.
By Claim 1, in both cases, we have $J_{\lambda_s,2}\subseteq A_{\diamond,s}\subseteq A^\dag$.\\
\noindent \textit{Step 2.} Consider the case $s=1$. Upon noting $\lambda_0 > \|\Psi^t y\|_{\infty}$,
there holds $J_{\lambda_1,3} = \emptyset$. To see this, we let $|x_i^\dag| = \max_{j=1,...,p}|x_j^\dag|$, then
\begin{equation*}
|\Psi_i^t y | \geq |x_i^\dag| - \delta\sqrt{T}|x_i^\dag| \geq \tfrac{3}{4}|x_i^\dag|\Rightarrow |x_i^\dag|<3\lambda_1 \Rightarrow J_{\lambda_1,3} = \varnothing.
\end{equation*}
By mathematical induction, noting the relations $\lambda_{s+1} = \frac{2}{3}\lambda_s$ and
$J_{\lambda_s,2} = J_{\lambda_{s+1},3}$, we conclude Claim 2.

For sufficient large $s$ s.t. $\lambda_0 \rho^s < \frac{1}{3}\min_{i\in A^\dag}|x_i^\dag|$,
then $J_{\lambda_s,3} = A^\dag$ and hence Algorithm \ref{PDAS} converges in one step and the
support of $x(\lambda_s)$ is $A^\dag$. The last assertion follows
\begin{equation*}
x^\dag - x(\lambda_s)_{A^\dag}  = (\Psi_{A^\dag}^t\Psi_{A^\dag})^{-1} d(\lambda_s)_{A^\dag}
\end{equation*}
and $\|d(\lambda_s)_{A^\dag}\|_{\infty} = \lambda_s$.
\end{IEEEproof}

\section*{Acknowledgment}
The work of Q. Fan was partially supported by National Science Foundation of China No. 61179039
and the work of X. Lu is partially supported by National Science Foundation of China No. 11101316
and No. 91230108. The authors would like to thank the anonymous referees for their constructive
 comments. The authors would also like to thank
Dr. Bangti Jin for useful discussions.

\ifCLASSOPTIONcaptionsoff
  \newpage
\fi

\bibliographystyle{IEEEtran}
\bibliography{IEEEabrv,jiaoyuling}

\begin{thebibliography}{10}
\providecommand{\url}[1]{#1}
\csname url@samestyle\endcsname
\providecommand{\newblock}{\relax}
\providecommand{\bibinfo}[2]{#2}
\providecommand{\BIBentrySTDinterwordspacing}{\spaceskip=0pt\relax}
\providecommand{\BIBentryALTinterwordstretchfactor}{4}
\providecommand{\BIBentryALTinterwordspacing}{\spaceskip=\fontdimen2\font plus
\BIBentryALTinterwordstretchfactor\fontdimen3\font minus
  \fontdimen4\font\relax}
\providecommand{\BIBforeignlanguage}[2]{{%
\expandafter\ifx\csname l@#1\endcsname\relax
\typeout{** WARNING: IEEEtran.bst: No hyphenation pattern has been}%
\typeout{** loaded for the language `#1'. Using the pattern for}%
\typeout{** the default language instead.}%
\else
\language=\csname l@#1\endcsname
\fi
#2}}
\providecommand{\BIBdecl}{\relax}
\BIBdecl

\bibitem{cs1}
E.~Cand\'{e}s, J.~Romberg, and T.~Tao, ``Robust uncertainty principles: Exact
  signal reconstruction from highly incomplete frequency information,''
  \emph{IEEE Transactions on Information Theory}, vol.~52, no.~2, pp. 489--509,
  2006.

\bibitem{cs2}
D.~Donoho, ``Compressed sensing,'' \emph{IEEE Transactions on Information
  Theory}, vol.~52, no.~4, pp. 1289--1306, 2006.

\bibitem{cs3}
J.~Tropp, ``Just relax: Convex programming methods for identifying sparse
  signals in noise,'' \emph{IEEE Transactions on Information Theory}, vol.~52,
  no.~3, pp. 1030--1051, 2006.

\bibitem{bp}
S.~Chen, D.~Donoho, and M.~Saunders, ``Atomic decomposition by basis pursuit,''
  \emph{SIAM Journal on Scientific Computing}, vol.~20, no.~1, pp. 33--61,
  1998.

\bibitem{lasso}
R.~Tibshirani, ``Regression shrinkage and selection via the lasso,''
  \emph{Journal of the Royal Statistical Society. Series B (Methodological)},
  pp. 267--288, 1996.

\bibitem{pareto}
E.~Van Den~Berg and M.~Friedlander, ``Probing the pareto frontier for basis
  pursuit solutions,'' \emph{SIAM Journal on Scientific Computing}, vol.~31,
  no.~2, pp. 890--912, 2008.

\bibitem{csnmr}
J.~Tropp and S.~Wright, ``Computational methods for sparse solution of linear
  inverse problems,'' \emph{Proceedings of the IEEE}, vol.~98, no.~6, pp.
  948--958, 2010.

\bibitem{BJMG}
F.~Bach, R.~Jenatton, J.~Mairal, and G.~Obozinski, ``Optimization with
  sparsity-inducing penalties,'' \emph{Foundations and Trends{\textregistered}
  in Signal Processing}, vol.~5, no. 1-2, 2012.

\bibitem{pfbsr}
P.~Combettes and J.~Pesquet, ``Proximal splitting methods in signal
  processing,'' \emph{Fixed-Point Algorithms for Inverse Problems in Science
  and Engineering}, pp. 185--212, 2011.

\bibitem{gpsr}
M.~Figueiredo, R.~Nowak, and S.~Wright, ``Gradient projection for sparse
  reconstruction: Application to compressed sensing and other inverse
  problems,'' \emph{IEEE Journal of Selected Topics in Signal Processing},
  vol.~1, no.~4, pp. 586--597, 2007.

\bibitem{sparsa}
S.~Wright, R.~Nowak, and M.~Figueiredo, ``Sparse reconstruction by separable
  approximation,'' \emph{IEEE Transactions on Signal Processing}, vol.~57,
  no.~7, pp. 2479--2493, 2009.

\bibitem{fpc}
E.~Hale, W.~Yin, and Y.~Zhang, ``Fixed-point continuation for
  $\ell_1$-minimization: Methodology and convergence,'' \emph{SIAM Journal on
  Optimization}, vol.~19, no.~3, pp. 1107--1130, 2008.

\bibitem{wen2010fast}
Z.~Wen, W.~Yin, D.~Goldfarb, and Y.~Zhang, ``A fast algorithm for sparse
  reconstruction based on shrinkage, subspace optimization, and continuation,''
  \emph{SIAM Journal on Scientific Computing}, vol.~32, no.~4, pp. 1832--1857,
  2010.

\bibitem{ista}
I.~Daubechies, M.~Defrise, and C.~De~Mol, ``An iterative thresholding algorithm
  for linear inverse problems with a sparsity constraint,''
  \emph{Communications on Pure and Applied Mathematics}, vol.~57, no.~11, pp.
  1413--1457, 2004.

\bibitem{pfbs}
P.~Combettes and V.~Wajs, ``Signal recovery by proximal forward-backward
  splitting,'' \emph{Multiscale modeling \& simulation}, vol.~4, no.~4, pp.
  1168--1200, 2005.

\bibitem{nest2}
Y.~Nesterov, ``Smooth minimization of non-smooth functions,''
  \emph{Mathematical programming}, vol. 103, no.~1, pp. 127--152, 2005.

\bibitem{fista}
A.~Beck and M.~Teboulle, ``A fast iterative shrinkage-thresholding algorithm
  for linear inverse problems,'' \emph{SIAM Journal on Imaging Sciences},
  vol.~2, no.~1, pp. 183--202, 2009.

\bibitem{nesta}
S.~Becker, J.~Bobin, and E.~Cand\'{e}s, ``{NESTA}: a fast and accurate
  first-order method for sparse recovery,'' \emph{SIAM Journal on Imaging
  Sciences}, vol.~4, no.~1, pp. 1--39, 2011.

\bibitem{homtop1}
M.~Osborne, B.~Presnell, and B.~Turlach, ``A new approach to variable selection
  in least squares problems,'' \emph{IMA journal of numerical analysis},
  vol.~20, no.~3, pp. 389--403, 2000.

\bibitem{homtop2}
B.~Efron, T.~Hastie, I.~Johnstone, and R.~Tibshirani, ``Least angle
  regression,'' \emph{The Annals of statistics}, vol.~32, no.~2, pp. 407--499,
  2004.

\bibitem{homtop3}
D.~Donoho and Y.~Tsaig, \emph{Fast solution of $l_1$-norm minimization problems
  when the solution may be sparse}.\hskip 1em plus 0.5em minus 0.4em\relax
  Department of Statistics, Stanford University, 2006.

\bibitem{admr}
S.~Boyd, N.~Parikh, E.~Chu, B.~Peleato, and J.~Eckstein, ``Distributed
  optimization and statistical learning via the alternating direction method of
  multipliers,'' \emph{Foundations and Trends in Machine Learning}, vol.~3,
  no.~1, pp. 1--122, 2011.

\bibitem{irls1}
I.~Daubechies, R.~DeVore, M.~Fornasier, and C.~Güntürk, ``Iteratively
  reweighted least squares minimization for sparse recovery,''
  \emph{Communications on Pure and Applied Mathematics}, vol.~63, no.~1, pp.
  1--38, 2009.

\bibitem{ssn1}
M.~Hinterm\"{u}ller, K.~Ito, and K.~Kunisch, ``The primal-dual active set
  strategy as a semismooth newton method,'' \emph{SIAM Journal on
  Optimization}, vol.~13, no.~3, pp. 865--888, 2002.

\bibitem{ssn3}
R.~Griesse and D.~Lorenz, ``A semismooth newton method for tikhonov functionals
  with sparsity constraints,'' \emph{Inverse Problem}, vol.~24, no.~3, pp.
  035\,007, 19 pp.

\bibitem{ssn4}
B.~Jin, D.~Lorenz, and S.~Schifler, ``Elastic-net regularization: error
  estimates and active set methods,'' \emph{Inverse Problem}, vol.~25, no.~11,
  pp. 115\,022, 26 pp., 2009.

\bibitem{orgth}
D.~Donoho and I.~Johnstone, ``Adapting to unknown smoothness via wavelet
  shrinkage,'' \emph{J. Amer. Statist. Assoc}, vol.~90, no. 432, pp.
  1200--1224, 1995.

\bibitem{roc}
R.~Rockafellar, \emph{Convex analysis}.\hskip 1em plus 0.5em minus 0.4em\relax
  Princeton university press, 1996, vol.~28.

\bibitem{general}
C.~Micchelli, L.~Shen, and X.~Y., ``Proximity algorithms for image models:
  denoising,'' \emph{Inverse Problems}, vol.~27, no.~05, pp. 045\,009, 30 pp.,
  2011.

\bibitem{ssn2}
K.~Ito and K.~Kunisch, \emph{Lagrange Multiplier Approach to Variational
  Problems and Applications}.\hskip 1em plus 0.5em minus 0.4em\relax SIAM,
  Philadelphia, 2008.

\bibitem{gloub}
G.~Golub and C.~Van~Loan, \emph{Matrix computations}.\hskip 1em plus 0.5em
  minus 0.4em\relax Johns Hopkins University Press, 1996, vol.~3.

\bibitem{SunQi2001}
D.~Sun and L.~Qi, ``Solving variational inequality problems via
  smoothing-nonsmooth reformulations,'' \emph{Journal of computational and
  applied mathematics}, vol. 129, no.~1, pp. 37--62, 2001.

\bibitem{HintermullerKunisch2006}
M.~Hinterm{\"u}ller and K.~Kunisch, ``Path-following methods for a class of
  constrained minimization problems in function space,'' \emph{SIAM Journal on
  Optimization}, vol.~17, no.~1, pp. 159--187, 2006.

\bibitem{Grasmair2011}
M.~Grasmair, O.~Scherzer, and M.~Haltmeier, ``Necessary and sufficient
  conditions for linear convergence of l1-regularization,''
  \emph{Communications on Pure and Applied Mathematics}, vol.~64, no.~2, pp.
  161--182, 2011.

\bibitem{Davenport2010}
M.~A. Davenport and M.~B. Wakin, ``Analysis of orthogonal matching pursuit
  using the restricted isometry property,'' \emph{Information Theory, IEEE
  Transactions on}, vol.~56, no.~9, pp. 4395--4401, 2010.

\bibitem{Engl1996}
H.~W. Engl, M.~Hanke, and A.~Neubauer, \emph{Regularization of inverse
  problems}.\hskip 1em plus 0.5em minus 0.4em\relax Springer, 1996, vol. 375.

\bibitem{Zhang2008}
C.-H. Zhang and J.~Huang, ``The sparsity and bias of the lasso selection in
  high-dimensional linear regression,'' \emph{The Annals of Statistics},
  vol.~36, no.~4, pp. 1567--1594, 2008.

\bibitem{bic2}
S.~Konishi and G.~Kitagawa, \emph{Information criteria and statistical
  modeling}.\hskip 1em plus 0.5em minus 0.4em\relax Springer, 2007.

\bibitem{bic5}
J.~Chen and Z.~Chen, ``Extended bayesian information criteria for model
  selection with large model spaces,'' \emph{Biometrika}, vol.~95, no.~3, pp.
  759--771, 2008.

\bibitem{df}
H.~Zou, T.~Hastie, and R.~Tibshirani, ``On the “degrees of freedom” of the
  lasso,'' \emph{The Annals of statistics}, vol.~35, no.~5, pp. 2173--2192,
  2007.

\bibitem{cosamp}
D.~Needell and J.~Tropp, ``Cosamp: Iterative signal recovery from incomplete
  and inaccurate samples,'' \emph{Applied and Computational Harmonic Analysis},
  vol.~26, no.~3, pp. 301--321, 2009.

\end{thebibliography}

\end{document}